\documentstyle[12pt,aaspp4]{article}
\begin{document}

\title{Transient Events from Neutron Star Mergers}

\author{Li-Xin Li and Bohdan Paczy\'nski}
\affil{Princeton University Observatory, Princeton, NJ 08544--1001, USA}
\affil{e-mail: lxl, bp@astro.princeton.edu}

\begin{abstract}

Mergers of neutron stars (NS+NS) or neutron stars and stellar mass black holes
(NS+BH) eject a small fraction of matter with a sub-relativistic velocity.
Upon rapid decompression nuclear density medium condenses into
neutron rich nuclei, 
most of them radioactive.  Radioactivity provides a long term heat source
for the expanding envelope.  A brief transient has the peak
luminosity in the supernova range, and the bulk of radiation in the
UV -- Optical domain.  We present a very crude model of the phenomenon,
and simple analytical formulae which may be used to estimate the parameters
of a transient as a function of poorly known input parameters.

The mergers may be detected with high redshift supernova searches
as rapid transients, many of them far away from the parent galaxies.
It is possible that the mysterious optical transients detected by
Schmidt et al. (1998) are related to neutron star mergers
as they typically have no visible host galaxy.

\end{abstract}

\keywords{gamma-rays: bursts -- stars: binaries: close -- 
stars: neutron -- stars: supernovae}

\section{Introduction}

Popular models of gamma-ray bursts (GRBs) include merging neutron stars 
(Paczy\'nski 1986, Popham et al. 1998, and references therein), and 
merging neutron stars and stellar mass black holes (Paczy\'nski 1991,
Popham et al. 1998, and references therein).  However, the location
of the recently detected GRB afterglows indicates that the bursts may
be located in star forming regions (Paczy\'nski 1998, Kulkarni et al. 
1998a,b, Taylor et al. 1998, Galama et al. 1998).  If this indication
is confirmed with the locations of the afterglows detected in the
future, then the NS+NS and NS+BH merger scenario will be excluded, as
those events are expected to occur far away from the place of their
origin (Tutukov \& Yungelson 1994, Bloom et al 1998, Zwart \& Yungelson 1998).

Still, the mergers are certainly happening, though at the rate estimated 
to be several orders of magnitude lower than supernova rate (Narayan
et al. 1991, Phinney 1991, van den Heuvel et al. 1996, 
Bloom et al 1998).  It is virtually
certain that a violent merger will eject some matter with a sub-relativistic
velocity.  The chemical composition of the ejecta must be very exotic
as it is formed by a rapid decompression of nuclear density matter.
It is not surprising that this process was suggested to be responsible
for some exotic elements (Lattimer \& Schramm 1974, 1976, Rosswog et al.
1998, and references therein).  As most nuclides are initially very neutron
rich, they will decay with various time scales.
Therefore, we expect a phenomenon somewhat similar
to a supernovae Type Ia, in which the decay of $ {\rm ^{56} Ni } $
first to $ {\rm ^{56}Co } $, and later $ {\rm ^{56}Fe } $ is responsible for 
the observed luminosity.  It is therefore interesting to explore the likely
light curves following the NS+NS and/or NS+BH mergers.

The neutron star mergers are expected to be among the first sources of
gravitational radiation to be detected by the LIGO (Abramovici et al. 1992).
It will be very important to detect the same events by other means.
In the last several years a lot of effort went to obtain gamma-ray bursts
from the mergers.  However, theoretical attempts are discouraging
(cf. Ruffert \& Janka 1997, 1998, and references therein), and the
observed locations of the burst afterglows does not favor the merger.
The purpose of this paper is to point out that the mergers are likely
to be accompanied with prominent optical transients, which should be 
detectable with the future supernova searches, and perhaps have already
been detected by Schmidt et al. (1998).

\section{Outline of the model}

Modeling of a NS+NS or a NS+BH merger is very complex, and our knowledge
of the outcome is very limited.  Therefore, instead of attempting to develop 
a complete numerical model (Ruffert \& Janka 1997, 1998, and references
therein) we make the simplest possible substitute,
a `one zone model' of an expanding envelope.  
While most of the mass falls into the black hole some matter is
ejected as a result of the complicated hydrodynamics of the merger,
or the powerful neutrino burst, or the super-strong magnetic fields.
For simplicity we assume that the expanding envelope is spherical,
its mass $ M $ is constant with time, and its density $\rho$ is uniform in
space and decreases with time.  The surface radius $R$ increases at the
fixed velocity $ V $, i.e. we ignore dynamical effect of the pressure 
gradient.  Therefore, the density throughout is given as
\begin{equation}
\rho = { 3 M \over 4 \pi R^3 } = \left( { 3 M \over 4 \pi V^3 } 
\right) t^{-3} ,
\label{rh}
\end{equation}
where $ t $ is the time from the beginning of expansion.  
All complications
of the initial conditions: the high temperature, the neutrino burst, and the
chemical composition, are absorbed into several input parameters of our
models: $ M $, $ V $, and the energy available for the radioactive decays.

The temperature inside the expanding sphere varies because of several
effects: adiabatic expansion, heat generation in radioactive decays, and 
radiative heat losses from the surface. Let us consider each of these 
effects.

{\em Adiabatic expansion.}  The density of expanding envelope rapidly becomes
very low, while the injection of a large amount of heat keeps it hot, and
radiation energy density dominates gas energy density.  Therefore, we adopt
\begin{equation}
U = 3 P = aT^4 ,
\label{u}
\end{equation}
where $U$ is the energy density, $P$ is the pressure, $T$ is the temperature,
and $a$ is the radiation constant.  Following the first law of thermodynamics 
the variation of entropy $S$ per unit mass is
\begin{equation}
T dS = { 1 \over \rho } ~dU + 
{ 4 \over 3 } U d \left( { 1 \over \rho } \right) 
\approx \left( { 4 \pi V^3 \over 3 M } \right) 
\left( t^3 dU + 4Ut^2 dt \right) .
\label{s}
\end{equation}

{\em Radioactive heating.} Assume that the radioactive decay of an
element isotopes proceeds on a time scale $t_{\rm rad}$, and releases a total
amount of energy equivalent to a fraction $f$ of the rest mass, so the heat
generation rate per gram per second is
\begin{equation}
\epsilon={fc^2\over t_{\rm rad}}\exp\left(-t/t_{\rm rad}\right).
\label{ep1}
\end{equation}
If there are several decaying element isotopes, the total heat generation
rate is the sum of that of individual element isotopes. If there are many
decaying element isotopes with different decaying time scales, the 
summation can be replaced by an integration. Nuclear lifetimes are
distributed roughly uniformly in logarithmic intervals in time, thus the 
total heat generation rate may be approximated as 
\footnote{We are very grateful to Dr. D. N. Spergel who suggested this 
formula to us.}
\begin{equation}
\epsilon={fc^2\over t} , \hskip 1.0cm
{\rm for} \hskip 1.0cm
t_{\rm min} \le t \le t_{\rm max} , \hskip 1.0cm
t_{\rm min} \ll t_{\rm max} .
\label{ep2}
\end{equation}

{\em Radiative losses.} The temperature gradient is $\sim T/R$, the average 
opacity is $\kappa\approx\kappa_{\rm e}\approx0.2 {\rm cm}^2{\rm g}^{-1}$
($\kappa_{\rm e}$ is the opacity caused by electron scattering), 
and the radiative
diffusion leads to the heat losses from the surface which are approximately
given by
\begin{equation}
F \equiv \sigma T_{\rm eff}^4 \approx {\sigma T^4 \over \kappa\rho R} , 
\hskip 1.0cm
L = 4 \pi R^2 F \approx \left( { 4 \pi ^2 V^4 c \over 3\kappa 
M } \right) U t^4 ,
\label{fl}
\end{equation}
where $\sigma = ac/4$ is the Stephan-Boltzmann constant, 
$T_{\rm eff}$ is the effective temperature. Throughout the paper
we adopt the diffusion approximation, which requires that the optical 
depth of the expanding sphere to be much larger than unity, i.e.
$\kappa\rho R\gg1$. The critical time $t_{\rm c}$ when the expanding sphere
becomes optically thin is given by $\kappa\rho R=1$, thus
\begin{equation}
t_{\rm c}=\left({3\kappa M\over 4\pi V^2}\right)^{1/2}
=1.13 ~ {\rm day}~\left({M\over 0.01M_\odot}\right)^{1/2}
\left({3V \over c}\right)^{-1}\left({\kappa\over\kappa_{\rm e}}\right)^{1/2}.
\end{equation}

The overall heat balance is given by
\begin{equation}
L \approx \left(\epsilon-T{dS\over dt}\right)M.
\label{l}
\end{equation}
Combining Eqs.~(\ref{rh}-\ref{l}), we obtain the equation for the variation
of internal energy
\begin{equation}
t^3 { dU \over dt } + 4 t^2 U \approx 
\left( { 3M \over 4 \pi V^3 } \right) \epsilon - 
\left( { \pi Vc \over\kappa M } \right) t^4 U .
\label{tem1}
\end{equation}

Define
\begin{eqnarray}
T_1&=&\left({4\pi f^2c^4\over 3a^2\kappa^3M}\right)^{1/8}
=2.80\times10^4~{\rm K}~\left(f\over 0.001\right)^{1/4}\left(
{M\over 0.01M_\odot}\right)^{-1/8}\left({\kappa\over\kappa_{\rm e}}
\right)^{-3/8},\nonumber\\
\tilde{U}&=&{U\over aT_1^4}, \hskip 1.0cm \tau={t\over t_{\rm c}},
\hskip 1.0cm \beta={V\over c},
\end{eqnarray}
then Eq.~(\ref{tem1}) can be written as
\begin{equation}
{d\tilde{U}\over d\tau}+\left({4\over\tau}+{3\tau\over4\beta}\right)\tilde{U}
={1\over\tau^3}~g(\tau),
\label{tem2}
\end{equation}
where
\begin{equation}
g(\tau)=\alpha e^{-\alpha\tau}, \hskip 1.0cm
\alpha={t_{\rm c}\over t_{\rm rad}}\approx 1.13
\left({M\over 0.01M_\odot}\right)^{1/2}\left({3V\over c}\right)^{-1}
\left({\kappa\over\kappa_{\rm e}}\right)^{1/2}
\left({t_{\rm rad}\over 1 ~ {\rm day}}\right)^{-1},
\label{g1}
\end{equation}
for the case of exponential law decay; or
\begin{equation}
g(\tau)={1\over\tau},
\label{g2}
\end{equation}
for the case of power law decay. 

Define
\begin{eqnarray}
L_0&=&{3fMc^2\over4\beta t_{\rm c}}=fc^3\left({3\pi M\over4\kappa}\right)^{1/2}
\nonumber\\
&=&4.13\times10^{44}~{\rm erg~s}^{-1}~\left({f\over 0.001}\right)
\left({M\over 0.01M_\odot}\right)^{1/2}
\left({\kappa\over\kappa_{\rm e}}\right)^{-1/2},
\end{eqnarray}
then the luminosity of the black body radiation can be written as
\begin{equation}
L=L_0\tau^4\tilde{U}.
\end{equation}
The effective temperature is given by
\begin{equation}
T_{\rm eff}=\tau^{1/2}T=T_1\tau^{1/2}\tilde{U}^{1/4}.
\end{equation}
All these equations are applicable to the optically thick case, i.e.
for $ \tau \le 1 $.

\section{Solutions of the model}

There are general properties of the Eq.~(\ref{tem2}) which do not
depend on the specific form of the radioactive energy generation
term, $ g( \tau ) $.  At the very beginning of expansion, at the
time of the order of a millisecond, the dimensionless time $ \tau \ll 1 $,
and the internal energy term $ \tilde{U} $ is likely to be very large.
The initial evolution is well approximated with the adiabatic expansion,
and the radiative heat loss term $ ( 3 \tau \tilde{U} / 4 \beta ) $,
and the heat generation term $ g( \tau )/ \tau ^3 $ can be both neglected.

The thermal evolution changes when the radioactive heat term becomes
important, and it changes again when the radiative heat losses from
the surface become important.  However, the initial phase of adiabatic
expansion makes the expanding sphere forget its initial thermal conditions.
For this reason, in all the subsequent discussion of the analytical
solution the initial conditions are found to be unimportant.

\subsection{Exponential law decay}
For the case
of exponential law decay, $g(\tau)$ is given in Eq.~(\ref{g1}). 
The analytic solution of Eq.~(\ref{tem2}) is
\begin{eqnarray}
\tilde{U}&=&{C_1\over\tau^4e^{3\tau^2/8\beta}}+
{4\alpha\beta e^{-\alpha\tau}\over3\tau^4}\times\nonumber\\
&&\times\left\{
1-\exp\left[-{3\over8\beta}\left(\tau-{4\over3}\alpha\beta\right)^2\right]
+\alpha\sqrt{8\beta\over3}~Y\left[\sqrt{3\over8\beta}\left(
\tau-{4\over3}\alpha\beta\right)\right]\right\},
\label{sol1}
\end{eqnarray}
where Dawson's integral $Y$ is defined by 
\begin{equation}
Y(x)=e^{-x^2}\int_0^xe^{s^2}ds=-Y(-x),
\end{equation}
$C_1$ is the integral constant which is determined by the initial condition.
The solution is practically insensitive to the initial conditions.
For example, for
$t_0\sim 1~{\rm ms}$, $T_0\sim10^{10}~{\rm K}$, $t_c\sim 10^5~{\rm s}$,
and $T_1\sim 10^4~{\rm K}$, the solution `forgets'
the initial conditions after $t=100~{\rm s}$.
Thus $C_1$ is roughly
\begin{eqnarray}
C_1\approx{4\over3}\alpha\beta
\left[-1+
\exp\left(-{2\over3}\alpha^2\beta\right)
+\alpha\sqrt{8\beta\over3}~Y\left(\alpha\sqrt{2\beta\over3}
\right)\right].
\end{eqnarray}
The luminosity is
\begin{eqnarray}
L&=&L_0 C_1e^{-3\tau^2/8\beta}+{4\over3}L_0 \alpha\beta e^{-\alpha\tau}
\times\nonumber\\
&&\times\left\{
1-\exp\left[-{3\over8\beta}\left(\tau-{4\over3}\alpha\beta\right)^2\right]
+\alpha\sqrt{8\beta\over3}~Y\left[\sqrt{3\over8\beta}\left(
\tau-{4\over3}\alpha\beta\right)\right]\right\}.
\label{lu11}
\end{eqnarray}

\begin{figure}[p]
\plotfiddle{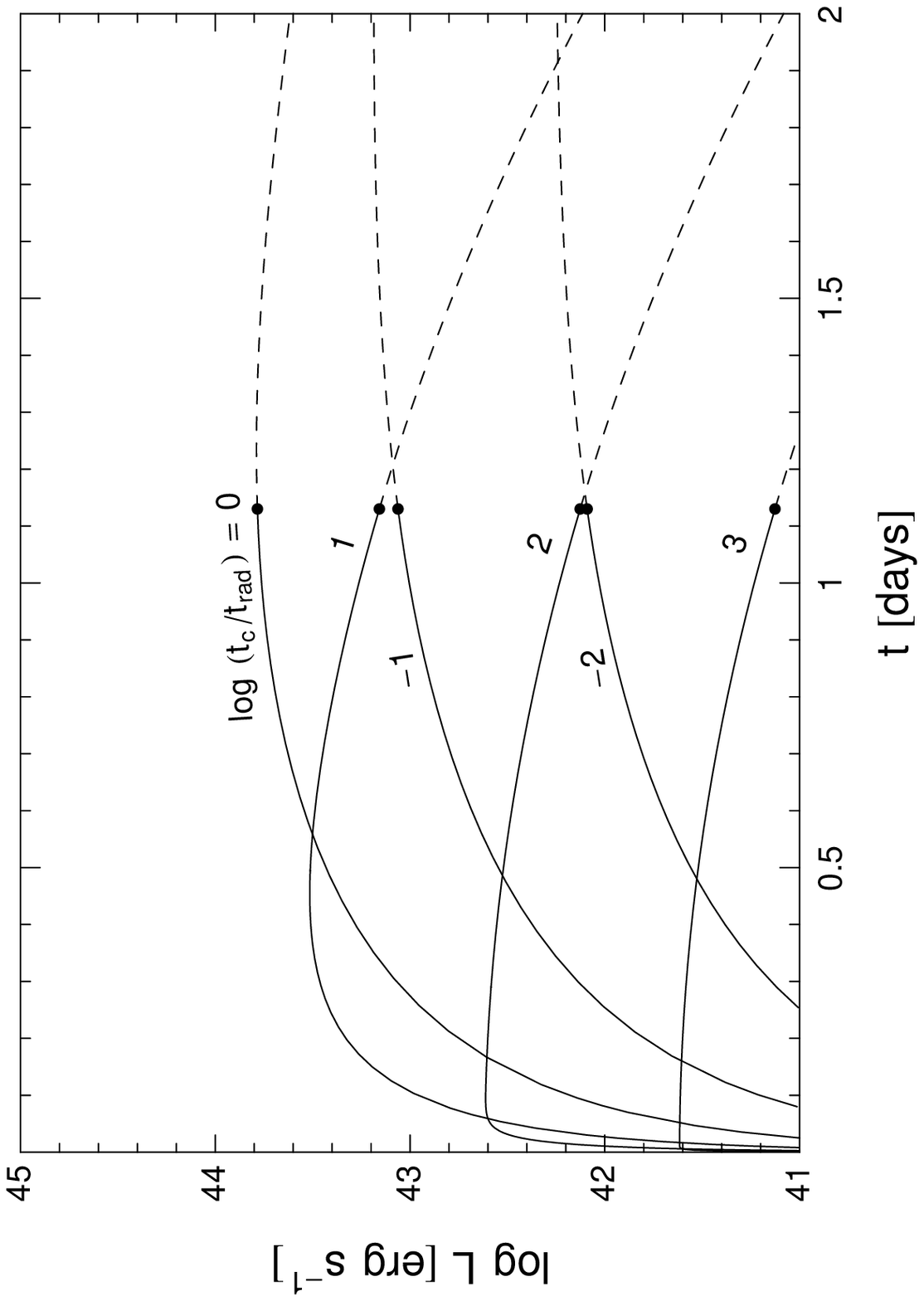}{6cm}{-90}{50}{50}{-200}{240}
\caption{ 
The time variation of the bolometric luminosity of the expanding
sphere generated by a neutron star merger is shown for a number of
models with various values of the logarithm of the ratio of two time scales:
$ t_{\rm c} $, when the sphere becomes optically thin, and the radioactive
decay time $ t_{\rm rad} $.  The models were calculated for 
the fraction of rest mass energy released in radioactive decay $ f = 10^{-3}$,
the mass $ M = 10^{-2}~ M_{\odot} $, and the surface expansion velocity 
$ V = 10^{10} {\rm cm ~ s^{-1} } $.  For the adopted opacity 
$ \kappa = 0.2 ~ {\rm cm^2 ~ g^{-1} } $ we have
$ t_{\rm c} = 0.975 \times 10^5 ~ {\rm s} = 1.13 ~ {\rm day} $,
as indicated with large dots, separating the continuous lines (corresponding
to the optically thick case) and the dashed lines (corresponding to the
optically thin case).
}
\plotfiddle{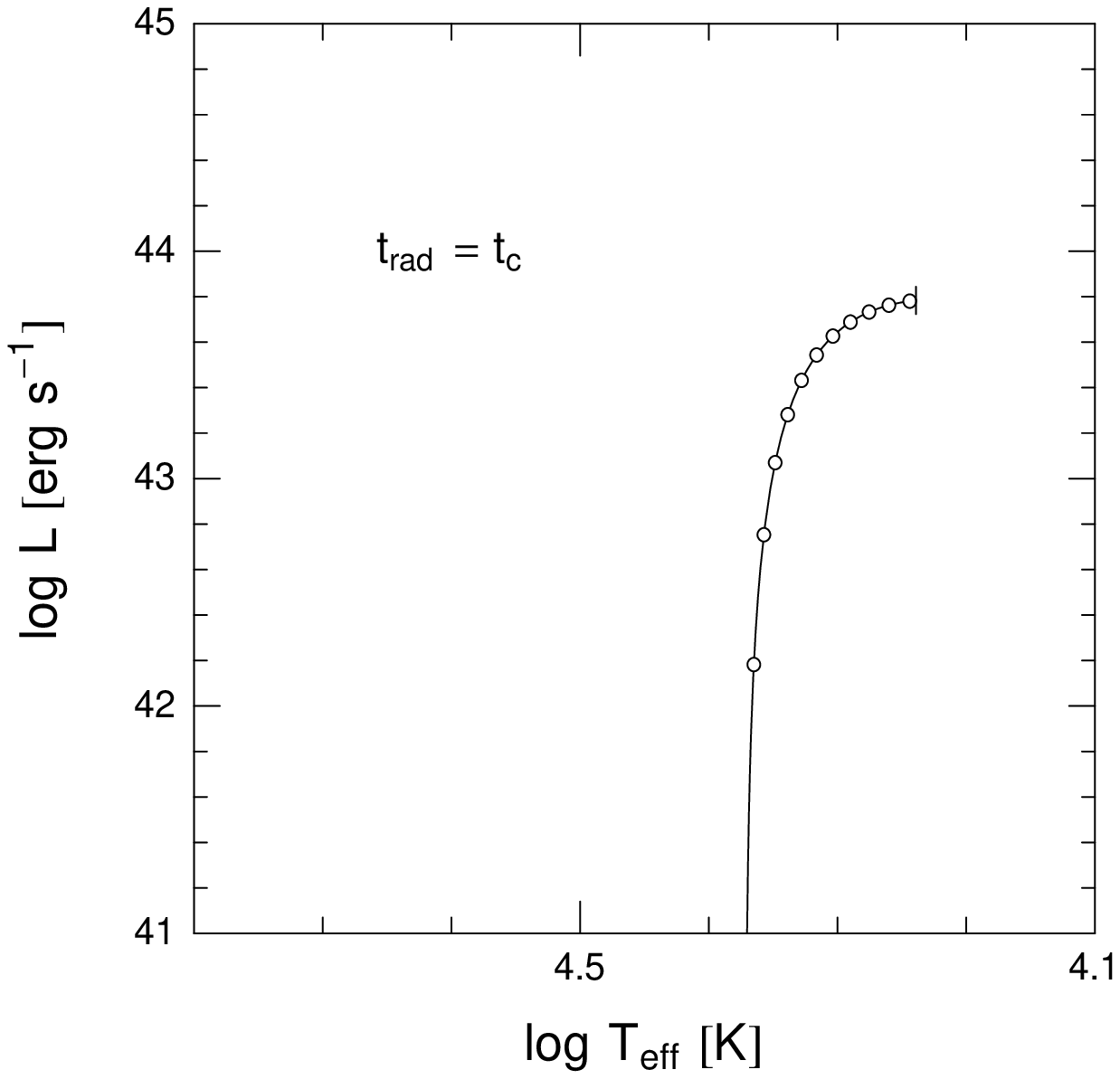}{6cm}{0}{50}{50}{-160}{-110}
\caption{
The evolution of the model from Fig.~1 with
the radioactive decay time $ t_{\rm rad} = 
t_{\rm c} = 0.975 \times 10^5 ~ {\rm s} = 1.13 ~ {\rm day} $
in the effective temperature -- bolometric luminosity diagram.  
The evolution begins at a very small
radius, with a very high $ T_{\rm eff} $ and a very low $ L $.  
The open circles correspond to the time of 0.1, 0.2, ... 1.1 days,
and the vertical bar corresponds to the time $ t_{\rm c} $ when the
model becomes optically thin.
}
\end{figure}

Though these exact solutions are interesting for theories, they are not
convenient to be used in practice. Thus here we also give some approximate
solutions which are simple and convenient to be used in practice and
have clear physical meanings.

We have already defined two time scales: $ t_{\rm rad} $, on which 
radioactive energy source decays exponentially, and $ t_{\rm c } $ on which
the optical depth of the expanding sphere is reduced to one.  There is
a third time scale $ t_{\rm d} \approx \kappa \rho R^2 /c = \beta t_c^2/t $
on which photons diffuse out of the sphere.  Of particular importance
is $ t_{\rm d,rad} \approx \alpha \beta t_{\rm c} $
when the photon diffusion time scale is equal to the radioactivity time
scale, and $ t_{\rm r} \approx \beta ^{1/2} t_{\rm c} $ when the
photon diffusion time becomes comparable to the expansion time.

If $\alpha^2\beta\gg1$ (thus $\alpha^{-1}\ll\beta^{1/2}
\ll\alpha\beta$), the evolution of the expanding sphere can be
divided into three regimes. The first regime is from $\tau\sim\tau_0$ to $\tau
\sim\beta^{1/2}$. In this regime
we may neglect the luminosity in the energy balance, i.e. we
may neglect the term $ ( 3 \tau \tilde{U} / 4 \beta ) $ in the
Eq.~(\ref{tem2}), and we obtain
\begin{equation}
\tilde{U} \approx { 1 \over \alpha \tau ^4 } 
\left[ 1 - ( 1 + \alpha \tau ) e^{ - \alpha \tau } \right] ,
\hskip 1cm
L \approx { L_0 \over \alpha } 
\left[ 1 - ( 1 + \alpha \tau ) e^{ - \alpha \tau } \right] ,
\hskip 1cm t \ll t_{\rm r} ,
\label{L2}
\end{equation}
$L$ reaches maximum at $ \tau _{\rm m} \approx 1 / \alpha $,
and $ t_{\rm m} \approx t_{\rm rad} $. 
The second regime is from $\tau\sim\beta^{1/2}$
to $\tau\sim\alpha\beta$. During this regime, since $\alpha\tau\gg 1$,
the radioactive term
$g(\tau)$ in the right hand side of Eq.~(\ref{tem2}) can be neglected, 
the resultant equation describes a radiation system cooling through
expanding and radiative processes. The solution of the resultant 
equation is
\begin{equation}
\tilde{U}\approx {1\over\alpha\tau^4}\exp\left(-{3\tau^2\over8\beta}\right),
\hskip 1cm
L\approx {L_0\over\alpha}\exp\left(-{3\tau^2\over8\beta}\right),
\label{sol4}
\end{equation}
where the integral constant is determined by joining this
solution with the solution before $\tau\sim\beta^{1/2}$. 
The third regime is for $\tau\gg\alpha\beta$, when
the photon diffusion time scale is much smaller than the
radioactive time scale. In such a case, the radiative energy loss is
balanced by the energy deposited from radioactive decay, and the 
luminosity is simply
\begin{equation}
L\approx\epsilon M
={4\over3}L_0\alpha\beta e^{-\alpha\tau}.
\label{sol7}
\end{equation}
Since $\alpha^2\beta
\gg1$, for the sub-relativistic case ($\beta\sim1$) this third regime
happens in the optically thin regime, the radiation given by
(\ref{sol7}) 
should be non-thermal. The peak luminosity is $L_{\rm m}\approx L_0/\alpha$,
which goes down as $\alpha$ increases.

If $\alpha^2\beta\ll1$ (thus $\alpha^{-1}\gg\beta^{1/2}\gg\alpha\beta$),
the evolution of the expanding shell can be divided into two regimes. The
first regime is from $\tau\sim\tau_0$ to $\tau\sim\beta^{1/2}$. During this 
period, since $\alpha\tau\ll 1$, the radioactive source $g(\tau)=
\alpha e^{-\alpha\tau}$ can be 
approximated by $g(\tau)\approx\alpha$. Inserting $g(\tau)\approx\alpha$
into Eq.~(\ref{tem2}), we get
\begin{equation}
\tilde{U}\approx{4\alpha\beta\over 3\tau^4}\left(
1-e^{-3\tau^2/8\beta}\right),
\hskip 1cm
L\approx {4\over3}L_0\alpha\beta\left(
1-e^{-3\tau^2/8\beta}\right).
\label{sol5}
\end{equation}
The second regime is for $\tau\gg\beta^{1/2}$, during which
the diffusion time scale is much smaller than the
radioactive time scale, the radioactive decay energy is balanced 
by radiative energy loss, and thus the luminosity is given by Eq.~(\ref
{sol7}). If it is extended into the optically thin regime, 
the radiation should become non-thermal for $\tau
\gg 1$. The peak of the luminosity
is roughly at the cross of the light curves given by Eqs.~(\ref{sol7})
and (\ref{sol5}), and the corresponding peak luminosity $ L_{\rm m} $
and time $ \tau_{\rm m} $ are
\begin{equation}
L_{\rm m}\approx {4\over3}\alpha\beta L_0 , \hskip 1cm
\tau_{\rm m}\approx\left[
-{4\beta\over3}\ln\left({16\alpha^2\beta\over3}\right)\right]^{1/2} . 
\end{equation}
The $ L_{\rm m} $ increases with $\alpha$. 

From the above discussion we see that, for $\alpha^2\beta\ll 1$,
the peak luminosity goes up as $\alpha$ increases; 
for $\alpha^2\beta\gg1$, the 
peak luminosity goes down as $\alpha$ increases. Thus, the peak luminosity
reaches its maximum roughly at $\alpha^2\beta\sim 1$
(or $t_{\rm rad}\sim\beta^{1/2}t_{\rm c}$, which is
comparable with $t_{\rm c}$ for sub-relativistic case), where we have
the biggest peak luminosity $L_m\sim \beta^{1/2}L_0$. 

The light curves are shown
in Fig.~1 for $M=10^{-2}M_\odot$, $V=c/3$, $f=10^{-3}$, $\kappa=0.2~
{\rm cm}^2{\rm g}^{-1}$ (thus $t_{\rm c}\approx0.975\times10^5$ s
$ \approx 1.13 $ day, $T_1\approx 2.8\times10^4$ K,
$L_0\approx4.1\times10^{44}$ erg/s, and $\beta=1/3$), and
$\alpha =t_{\rm c}/t_{\rm rad}=
10^{3}$, $10^{2}$, $10$, $1$, $10^{-1}$, and $10^{-2}$ respectively. 
The initial
condition is taken to be $T(t=1~{\rm ms})=2.8\times 10^{10}~{\rm K}$ (but the
results are insensitive to the initial condition).
Since for 
$ \left( 3/8\beta \right) ^{1/2} \left( \tau- 4 \alpha\beta /3 \right)\gg1$
we have $L\approx {4\over3}L_0\alpha\beta e^{-\alpha\tau}$ which
joins smoothly with $L\approx \epsilon M$, Eq.~(\ref{lu11})
can be formally extended to optically thin region, but 
we must keep in mind that in optically thin case the radiation
is non-thermal.

The evolution of our model with the exponential decay law
is shown in the $ \log T_{\rm eff} - \log L $ diagram in Fig.~2 
for the case of $t_{\rm rad}=t_{\rm c}$. 

The numerical results agree very well with our
analytical estimates of the parameters important for observers:
the time from the beginning of expansion to the peak luminosity, 
the peak luminosity, the temperature at the peak luminosity, and 
the time it takes for the luminosity to fall down by a factor 3 from 
the peak.

\subsection{Power law decay}

The exponential decay model was useful in demonstrating that the very
short and very long time scale radioactivity is of little use for
generating a large luminosity.  The most efficient conversion of
nuclear energy to the observable luminosity is provided by the elements
with the decay time scale $ t_{\rm rad} $ comparable to $ t_{\rm c} $, 
when the expanding sphere becomes optically thin.  In reality, there is 
likely to be a large number of nuclides with a very broad range of decay 
time scales.  Therefore, it is more realistic to adopt a power law decay 
model, which automatically selects the most efficient radioactive time scales.

In the case of power law decay, $g(\tau)$ is given with Eq.~(\ref{g2}).
The analytic solution of Eq.~(\ref{tem2}) is
\begin{equation}
\tilde{U}={C_4\over\tau^4e^{3\tau^2/8\beta}}+
\sqrt{8\beta\over3}~{1\over\tau^4}~Y\left(\sqrt{3\over8\beta}~\tau\right),
\label{sol2}
\end{equation}
where the integration constant $C_4$ is determined by the initial condition.
Similar to the case of exponential law decay, in practice
the solution is insensitive to the initial condition. In this
approximation, we have $C_4\approx 0$.
The corresponding luminosity is given by 
\begin{equation}
L\approx L_0\sqrt{8\beta\over3}~Y\left(
\sqrt{3\over 8\beta}~\tau\right).
\label{lu2}
\end{equation}

\begin{figure}[p]
\plotfiddle{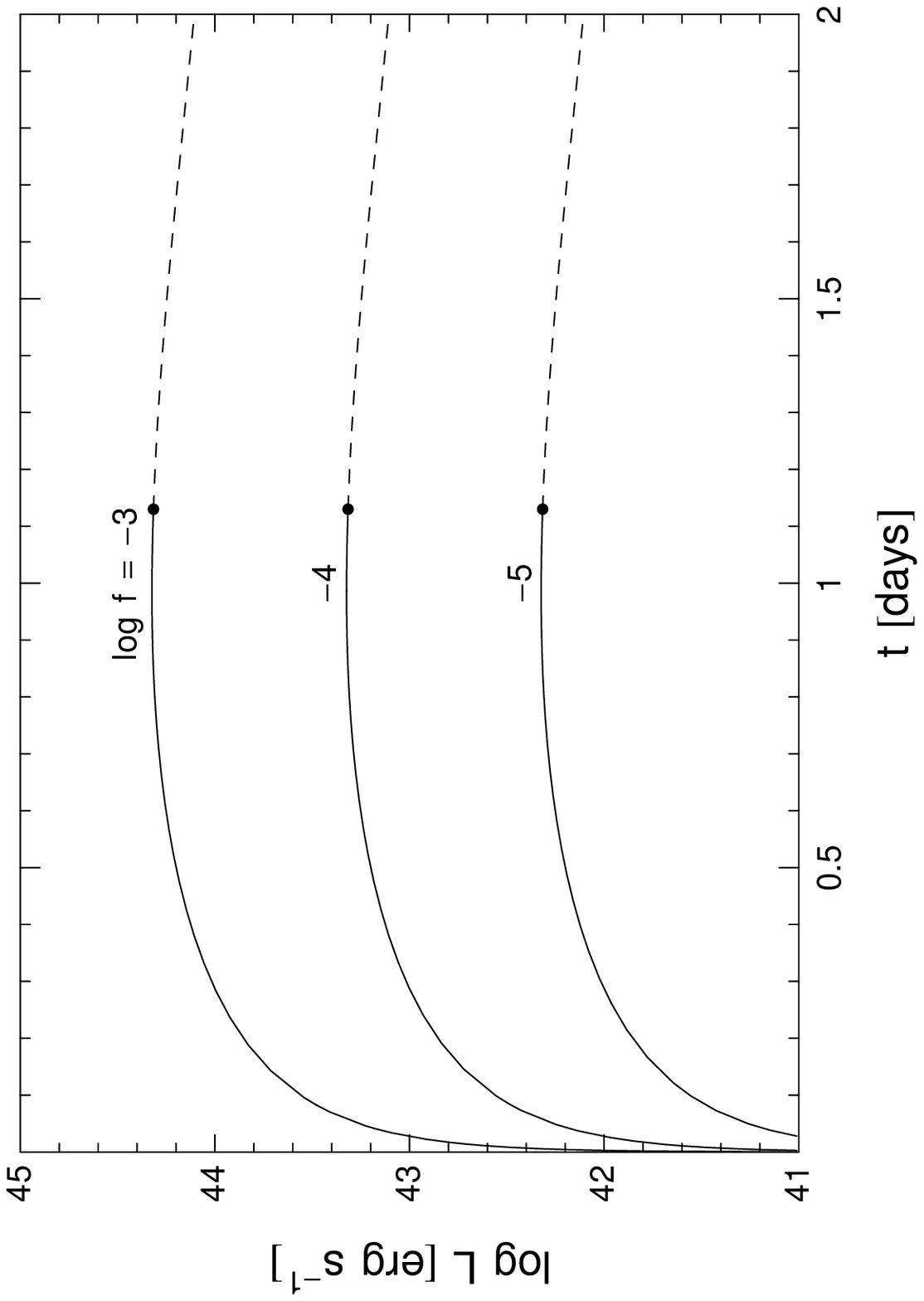}{6cm}{-90}{50}{50}{-200}{240}
\caption{ 
The time variation of the bolometric luminosity of the expanding
sphere generated by a neutron star merger is shown for
models with a large mix of radioactive nuclides which provide 
a heating rate inversely proportional to time from the beginning
of expansion.  The models have three values of the fraction of rest mass 
released as heat: $ f = 10^{-3}, ~ 10^{-4}, ~ 10^{-5} $,
the mass $ M = 10^{-2}~ M_{\odot} $, and the surface expansion velocity 
$ V = 10^{10} {\rm cm ~ s^{-1} } $. For the adopted opacity 
$ \kappa = 0.2 ~ {\rm cm^2 ~ g^{-1} } $ we have 
$ t_{\rm c} = 0.975 \times 10^5 ~ {\rm s} = 1.13 ~ {\rm day} $, as
indicated with large dots, separating the continuous lines (corresponding
to the optically thick case) and the dashed line (corresponding to the
optically thin case). 
}
\plotfiddle{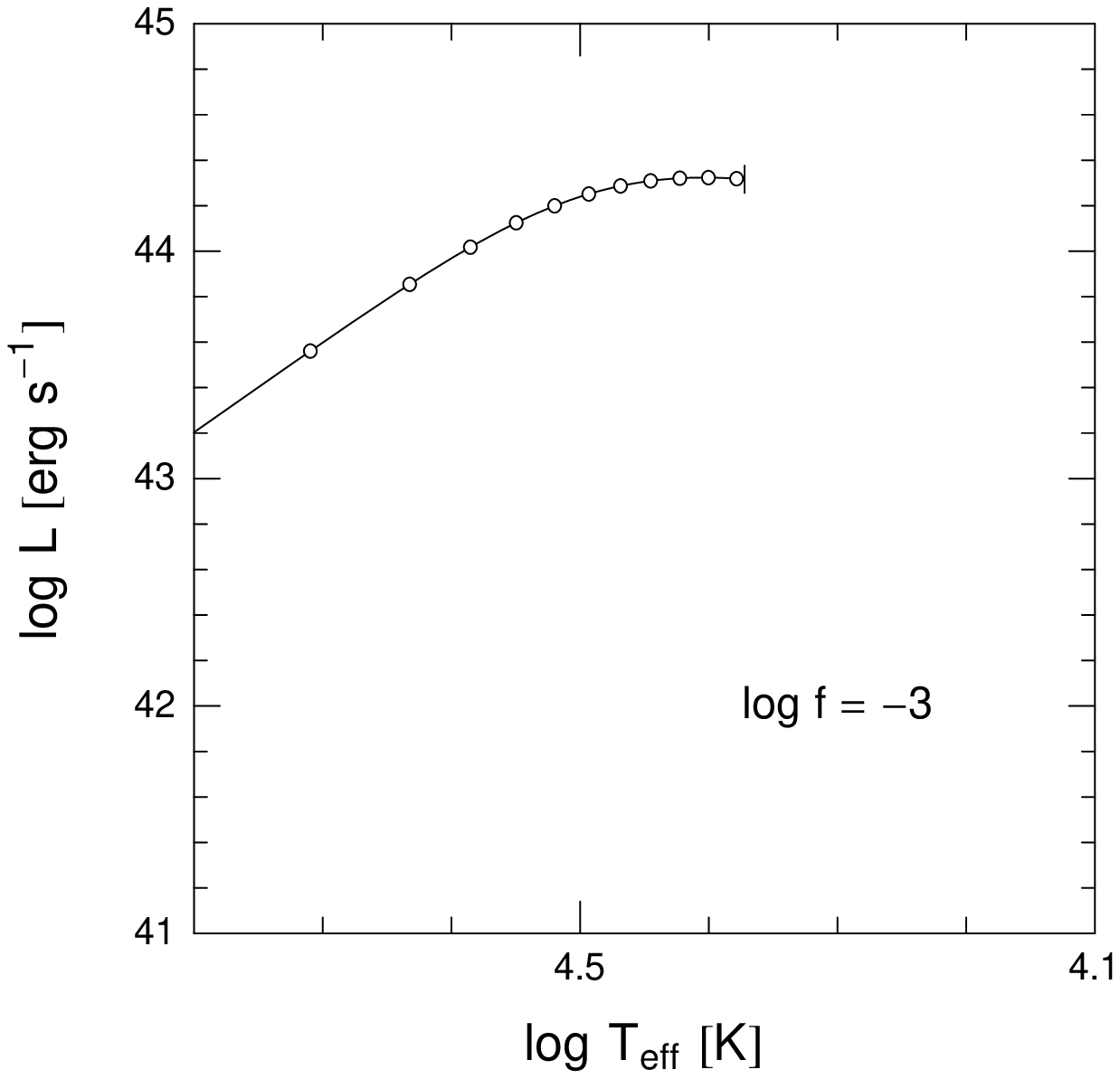}{6cm}{0}{50}{50}{-160}{-110}
\caption{
The evolution of the model from Fig.~3 with
the fraction of rest mass energy released 
in radioactive decay $ f = 10^{-3} $.
The evolution begins at a very small
radius, with a very high $ T_{\rm eff} $ and a very low $ L $.  
The open circles correspond to the time of 0.1, 0.2, ... 1.1 days,
and the vertical bar corresponds to the time
$ t_{\rm c} = 0.975 \times 10^5 ~ {\rm s} = 1.13 ~ {\rm day} $ when the
model becomes optically thin.
}
\end{figure}

For positive $x$, $Y(x)$ has the maximum value at $x_{\rm m}\approx0.9241$ and 
$Y_{\rm m}\approx0.5410$. Thus,
the time from the beginning of expansion to the peak luminosity is
\begin{equation}
t_{\rm m}\approx 1.5\beta^{1/2}t_{\rm c}= 0.98 ~ {\rm day}~
\left({M\over 0.01M_\odot}\right)^{1/2}\left({3V\over c}\right)^{-1/2}
\left({\kappa\over\kappa_{\rm e}}\right)^{1/2}.
\label{lu3}
\end{equation}
The peak luminosity is
\begin{equation}
L_{\rm m}\approx 0.88\beta^{1/2}L_0=2.1\times10^{44}~{\rm erg~s}^{-1}~
\left({f\over0.001}\right)
\left({M\over 0.01M_\odot}\right)^{1/2}\left({3V\over c}\right)^{1/2}
\left({\kappa\over\kappa_{\rm e}}\right)^{-1/2}.
\label{lu4}
\end{equation}
The effective temperature at the peak luminosity is
\begin{equation}
T_{\rm eff,m}\approx0.79\beta^{-1/8}T_1=2.5\times10^4{\rm K}~
\left({f\over0.001}\right)^{1/4}
\left({M\over 0.01M_\odot}\right)^{-1/8}\left({3V\over c}\right)^{-1/8}
\left({\kappa\over\kappa_{\rm e}}\right)^{-3/8}.
\label{lu5}
\end{equation}
From Eq.~(\ref{lu2}), the time when the luminosity is down a factor 3 
from the peak luminosity is 
$t\approx4.9\beta^{1/2}t_{\rm c}$, which is in
the optically thin epoch for the sub-relativistic case ($V\sim0.1c - 0.5c$).
For the optically thin case, the luminosity of the 
(non-thermal) radiation from the expanding shell is roughly given by
$L\approx\epsilon M\approx {4\over3}
L_0\beta\tau^{-1}$, which is formally the 
asymptotic form of Eq.~(\ref{lu2}) for $\tau\gg\beta^{1/2}$
[$Y(x)\approx(2x)^{-1}$ for $x^2\gg1$]. Thus, Eq.~(\ref{lu2}) can be
formally extended to the optically thin region, where the radiation
is non-thermal.  The time from the peak 
luminosity to the luminosity down a factor 3 from the peak is roughly
given by
\begin{equation}
\Delta t\approx 3.4\beta^{1/2}t_{\rm c}= 2.2 ~ {\rm day}~
\left({M\over 0.01M_\odot}\right)^{1/2}\left({3V\over c}\right)^{-1/2}
\left({\kappa\over\kappa_{\rm e}}\right)^{1/2}.
\label{lu6}
\end{equation}

Fig.~3 shows light curves drawn from Eq.~(\ref{lu2}) with $M=10^{-2}M_\odot$,
$V=c/3$, $\kappa=0.2~{\rm cm}^2{\rm g}^{-1}$, 
and $f=10^{-3}$, $10^{-4}$, $10^{-5}$
respectively. Fig.~4 shows the evolution of the expanding shell in the
$ \log T_{\rm eff} - \log L $ diagram for the case of $f=10^{-3}$. 
Note that the peak luminosity is proportional to $ f $, and the shape of
the light curve, and all the time scales, are independent of $ f $ [cf. 
Eqs.~(\ref{lu2} - \ref{lu6})].

\section{Conclusion} 

Our model is so simple, that it can provide only an order of magnitude
estimate of the peak luminosity and the time scale of a transient event
that is likely to follow a violent merger of two neutron stars or
a merger between a neutron star and a stellar mass black hole.  The 
Eqs.~(\ref{lu2} - \ref{lu6}), and Figs.~3 and 4 provide a convenient 
representation
of our model, and a simple relation between the poorly known input 
parameters: the mass of the ejecta $ M $, their velocity $ V $, and the
fraction of rest mass energy available for radioactive decays, and the
observable parameters.  Note, that for a plausible set of input parameters
the transient reaches peak luminosity $ L_{\rm m} \approx 10^{44} ~ {\rm erg ~
s^{-1} } $, corresponding to the bolometric luminosity of $ M_{\rm bol} \approx
-21 $, i.e. in the bright supernova range.  However, the duration of the
luminous phase is likely to be only $ t_{\rm m} \approx 1 $ day, i.e.
much shorter than a supernova.  The duration can be extended if the
ejecta have a large mass and expand slowly [cf. Eq.~(\ref{lu3})].

While there are many improvements our model requires, the single most
important is a quantitative estimate of the abundances and the lifetimes
of the radioactive nuclides that form in the rapid decompression of
nuclear density matter.  It is possible that Rosswog et al. (1998)
may readily provide this improvement.

As the frequency of such events is expected to be $ \sim 10^3 $ times
lower than the supernova rate
(Narayan et al. 1991, Phinney 1991, van den Heuvel et al. 1996, 
Bloom et al 1998), they may be detected soon in the 
supernovae searches.  The merger events are likely to be hotter
than supernovae, which can make them easier to detect
at large redshifts.  The dust extinction which affects some supernovae
is not likely to be a problem, as the mergers are expected far from
star forming regions, many of them (perhaps most of them) outside
of parent galaxies (Tutukov \& Yungelson 1994, Bloom et al 1998, 
Zwart \& Yungelson 1998). 

It is very intriguing that the high redshift supernova search (Schmidt at
al. 1998) revealed mystery optical transients which typically have no host
galaxy, as would be expected of neutron star mergers. In principle it
should be fairly straightforward to test our suggestion by measuring
future mystery events in at least four photometric bands, so that at
least three colors would be available. In the optically thick phase our
model would follow a locus occupied by ordinary stars in the
color-color-color diagram, evolving along a line corresponding
to a fixed redshift and changing effective temperature. As soon as
the expanding matter becomes optically thin the colors would deviate
significantly from the area occupied by any star with a more or less normal
photosphere. 

\acknowledgments{It is a great pleasure to thank Dr. D. N. Spergel for
asking us a question: `what might be the observable consequences of neutron
star mergers?', and for his useful comments.
One of us (BP) acknowledges hospitality he experienced
while on his sabbatical at the Institut d'Astrophysique, CNRS in Paris.}


\end{document}